\documentclass[aps,prl,twocolumn,showpacs]{revtex4}
\usepackage{graphicx}
\usepackage{dcolumn}
\usepackage{bm}
\usepackage{epsfig}
\usepackage{amsmath}
\usepackage{amssymb}
\usepackage{bm}
\usepackage{amstext}
\usepackage{amsfonts}
\usepackage{amssymb}
\usepackage{mathrsfs}

\newcommand{\ket}[1]{|#1\rangle}
\newcommand{\bra}[1]{\langle#1|}
\newcommand{\braket}[2]{\langle#1|#2\rangle}

\newcommand{\ncd}{\newcommand}
\ncd{\QC}{$\mbox{QC}$}
\ncd{\QCpr}{${\mbox{QC}_{\cal{C}}}^\prime\;$}
\ncd{\QCns}{$\mbox{QC}_{\cal{C}}$}
\ncd{\QCprns}{${\mbox{QC}_{\cal{C}}}^\prime$}
\ncd{\cskN}{{|\phi_{\{\kappa\} } \rangle}_{{\cal{C}}_N}}
\ncd{\cskNpr}{{|\phi_{\{\kappa^\prime\} } \rangle}_{{\cal{C}}_N}}
\ncd{\cskNtil}{{|\phi_{\{\tilde{\kappa} \} }
\rangle}_{{\cal{C}}_N}}
\ncd{\csk}{{|\phi_{\{\kappa\} }
\rangle}_{\cal{C}}}
\ncd{\csktil}{{|\phi_{\{\tilde{\kappa} \} }
\rangle}_{\cal{C}}}
\ncd{\cskf}{|\phi_{\{\kappa\} }
\rangle_{\cal{C}}}
\ncd{\csktilf}{|\phi_{\{\tilde{\kappa} \} }
\rangle_{\cal{C}}}
\ncd{\bracsk}{\mbox{}_{\cal{C}}\langle\phi_{\{\kappa\} }|}
\ncd{\bracsktil}{\mbox{}_{\cal{C}}\langle\phi_{\{\tilde{\kappa} \}
}|} \ncd{\nbracsk}{\mbox{}_{\cal{C}}\langle\phi_{\{\kappa\} }}
\ncd{\nbracsktil}{\mbox{}_{\cal{C}}\langle\phi_{\{\tilde{\kappa}
\} }} \ncd{\cs}{|\phi \rangle_{\cal{C}}\;} \ncd{\csns}{|\phi
\rangle_{\cal{C}}}
\ncd{\nbgh}{\text{nbgh}} 
\ncd{\SABC}{S^{ABC}}
\ncd{\Sab}{S^{ab}}
\ncd{\Sbc}{S^{bc}}
\ncd{\Sba}{S^{ba}}
\ncd{\ds}{\displaystyle} \ncd{\ovl}{\overline}

\begin{document}
\setlength{\hoffset}{-0.2 cm}

\title{Natural three-qubit interactions in one-way quantum computing}
\author{M. S. Tame$^1$, M. Paternostro$^1$, M. S. Kim$^1$, V. Vedral$^2$}
\affiliation{$^1$School of Mathematics and Physics, The Queen's University, Belfast, BT7 1NN, UK\\
$^2$The School of Physics and Astronomy, University of Leeds, Leeds, LS2 9JT, UK}
\date{\today}
\begin{abstract}
We address the effects of natural three-qubit interactions on the computational power of one-way quantum computation (\QC).  
A benefit of using more sophisticated entanglement structures is the ability to construct compact and economic simulations of 
quantum algorithms with limited resources.
We show that the features of our study are embodied by suitably prepared optical lattices, where effective three-spin interactions have been theoretically demonstrated. We use this to 
provide a compact construction for the
Toffoli gate. Information flow and two-qubit interactions are also outlined, together with a brief analysis of relevant sources of imperfection.

\end{abstract}

\maketitle

Measurement based one-way quantum computation (\QC)~\cite{RH,RBH} is an intriguing alternative to the standard quantum circuit model and has recently generated enormous interest for its application to quantum information processing (QIP). From the one-way perspective, the simulation of quantum gates is performed by adaptive single-qubit measurements on suitably prepared multipartite entangled resources known as {\it cluster-states}. These constitute a particular class of the more general graph-states~\cite{HEB} whose properties have been extensively studied~\cite{DH}. Very recently, the basic features of cluster-state based \QC~and a two-qubit quantum search algorithm have been experimentally demonstrated~\cite{Wal,Pan}.

However, the realization of cluster-state based quantum algorithms is often expensive in terms of qubit resources, an aspect quite detrimental to the efficiency of the quantum computation~\cite{Tame1,Tame2}. An illuminating example is provided by the {$n$-qubit generalization} of the simple two-qubit {\it searching for a marked entry} algorithm realized in~\cite{Wal}.
In the standard network model for \QC~\cite{Bar} this consists of ${\cal O}(\sqrt{2^n})$ oracle-inversion steps~\cite{Grov} each requiring two {$n$-time controlled-$\sf NOT$ ({\sf CNOT}) gates}.
For $n>3$, these can be made from $4(n-3)$ three-qubit Toffoli gates  ({\sf C$^2$NOT})~\cite{Bar}, which in a cluster-state based implementation require $65$ qubits~\cite{RBH}. 
A {three}-qubit version of the algorithm therefore requires $\sim 245$ cluster qubits~\cite{Conc}, a number which makes the protocol susceptible to even small amounts of noise affecting the cluster resource. A way to counteract this difficulty is the use of more compact designed cluster configurations which simulate three-qubit gates. 
If the universal three-qubit Toffoli gate is realized in a compact way, the number of qubits and manipulations needed to perform a given task will be dramatically reduced.  
Unfortunately any attempt in this direction in a cluster-state based scenario seems to be destined to failure.
The reason is due to the underlying two-qubit structure imposed by the effective control-phase ({\sf CP}) gates used {in the construction of} cluster-states~\cite{RH}: a constraint preventing any compact natural three-qubit gate is set. It is therefore interesting to investigate whether other entanglement structures are possible for the multipartite entangled resource, providing economical configurations which scale better in the presence of noise.
In this work we describe one such possibility, based on recently demonstrated three-spin interactions in optical lattices~\cite{Pachos1}. 
We first introduce the entanglement structure of the resource and comment on {simulations of} QIP protocols via measurements.
Next we provide a physically realizable 
setup {for our proposal},
in the form of a {\it bowtie} shaped optical superlattice. Finally, imperfections within the model at the entanglement stages {are} briefly addressed. 
Our proposal allows the construction of compact configurations for the simulation of Toffoli gates in one-way \QC~and opens up new possibilities in the search for conducting economical and robust-to-noise quantum algorithms.

\begin{figure}[t]
\centerline{{\bf (a)}\hskip3.5cm{\bf (b)}}
\centerline{\psfig{figure=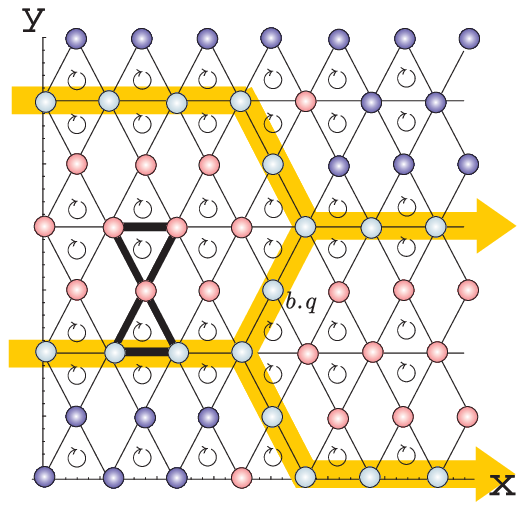,width=3.8cm,height=3.6cm}\psfig{figure=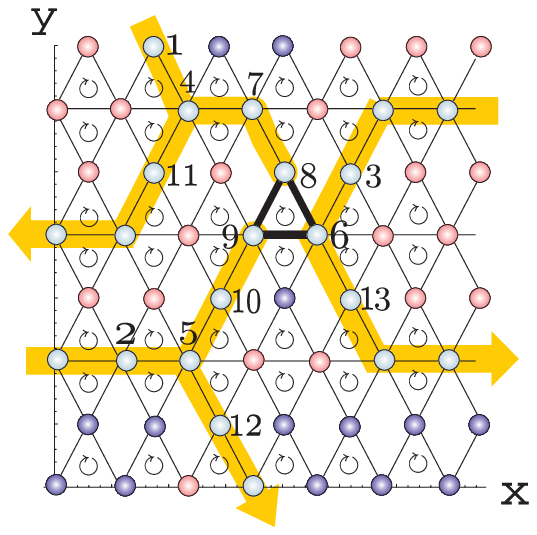,width=3.8cm,height=3.6cm}}
\caption{{\bf(a)}: The lattice structure and propagation of quantum information. Red (blue) dots represent qubits in $\ket{1}$ ($\ket{0}$) and $\circlearrowright$ denotes the three-spin interaction. The dark arrow shows the path of information, which is propagated via $\sigma_{x}$ measurements. A two qubit interaction is also shown, where the {\it bridging} qubit ($b.q$) should be measured in the $\sigma_y$ eigenbasis \cite{Tame2}. {\bf (b)}: A three-qubit interaction and the paths to bring information in/out of the interaction region.}
\label{threespin1}
\end{figure}

{\it The model}- We consider a lattice of qubits with the {\it bowtie} structure depicted in Fig.~\ref{threespin1}~{\bf (a)}, where each qubit with logical basis $\{\ket{0},\ket{1}\}$ is initially prepared in the state $\ket{+}=(1/\sqrt{2})(\ket{0}+\ket{1})$.
For every closed triangle, an entangling operation is applied between qubits $i,~j$ and $k$ at the vertices equivalent to a control-control-phase gate (${\sf C^2P}$),
$S^{ijk}={\openone}^{(ijk)}-2|111 \rangle_{ijk} \langle 111|$.
For convenience we denote this operation by $\circlearrowright$ and use $\sigma_{l,i}\,(l=x,y,z)$ as the $l$-Pauli matrix applied to the $i$-th qubit. The physical mechanism which realizes this configuration is addressed later. We first describe a way to realize information-flow across the lattice. In order to create a path for information to be propagated along via measurements, it is necessary to remove the influence of particular lattice qubits depending on the protocol being performed. 
It is straightforward to check that due to the three-body nature of the entanglement, a measurement in the single-qubit $\sigma_z$ eigenbasis with outcome $\ket{0}$ ($\ket{1}$) destroys (sustains) entanglement between the remaining two qubits. On the other hand, by setting the qubit to be removed in $\ket{0}$ or $\ket{1}$ before entanglement is generated across the entire lattice, a path can be formed, the choice being dependent on the required shape of the path itself. Setting a qubit to $\ket{1}$ generates the well-known Ising-type interaction between the {other} two, 
while setting it to $\ket{0}$ prevents any interaction from being generated. Paths of linear cluster-states may then be embedded within the lattice and used to propagate information using techniques borrowed from the cluster-state model~\cite{RBH}, as shown in Fig.~\ref{threespin1} {\bf (a)}. This technique, which initializes the qubits not involved in a specific protocol before introducing the entanglement across the lattice, 
puts the {\it removed} qubits in an eigenstate of $\sigma_z$. This vastly reduces the effects of spreading measurement or environment-induced noise created at the entangling stage~\cite{Tame1}.
It is easy to see that two-qubit gates can be realized in a similar way to the cluster-state model. An example is given in Fig.~\ref{threespin1} {\bf (a)}, where measuring the {\it bridging qubit} $(b.q)$ in the $\sigma_y$ eigenbasis simulates the gate $U={\sf CNOT}(\openone \otimes R_z^{\pi / 2}) {\sf CNOT}$ \cite{Tame2}, with $R_z^{\pi / 2}$ a single-qubit rotation about ${\bf \hat{z}}$ on the Bloch sphere by $\pi / 2$.

\begin{figure}[t]
\hskip-0.3cm{\bf (a)}\hskip3.5cm{\bf (b)}
\centerline{\psfig{figure=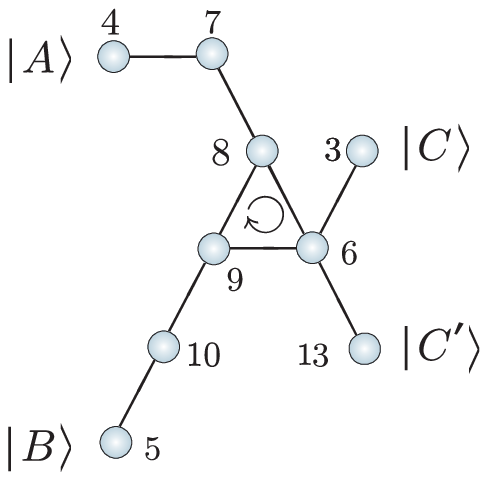,width=3.0cm,height=2.8cm}\hskip1.0cm\psfig{figure=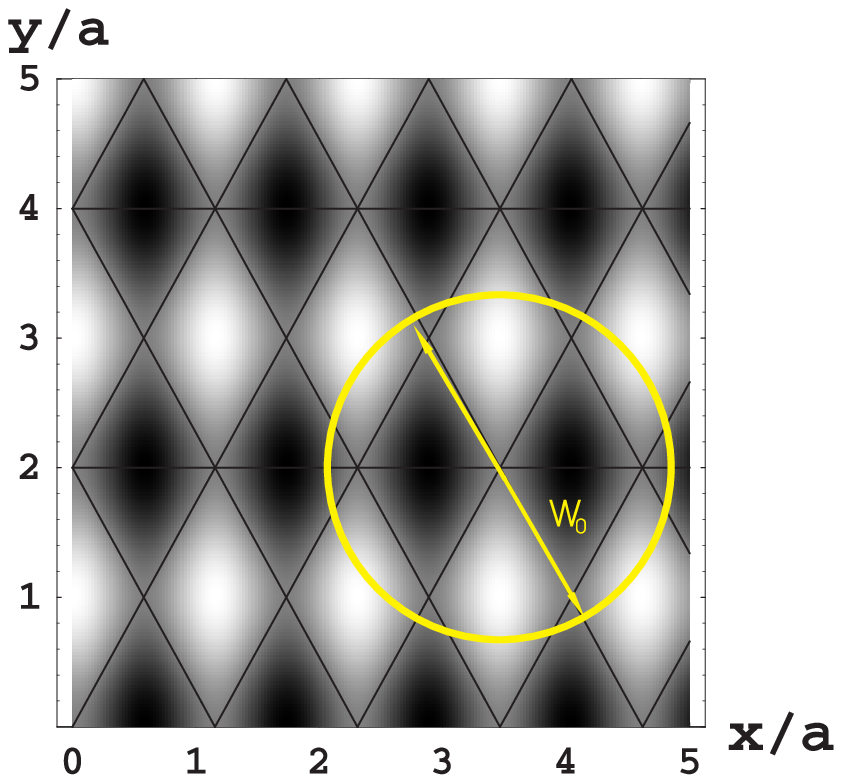,width=3.8cm,height=3.6cm}}
\caption{
{\bf (a)}: A ${\sf C^2 NOT}$ extracted from Fig.~\ref{threespin1} {\bf (b)}. Qubit $\ket{C}$ is the target and $\circlearrowright$ naturally realizes a $\sf C^2P$. 
$\sigma_{x}$-measurements of qubits $3$ and $6$ realize Hadamard gates before and after $\circlearrowright$. {\bf (b)}: The {\it bowtie} lattice structure created by $V_{off}$, where the $x$ and $y$ axes are scaled by $a=\lambda/2$. Dark (light) regions correspond to small (large) positive valued potential shifts, with the vertices of the superimposed grid representing lattice sites. The width  $w_0\sim2.8\,a$ of a Gaussian beam (used for register initialization) is shown.} 
\label{threespin2}
\end{figure}

In addition to the embedded standard cluster-state based manipulation of information, the three-spin entanglement structure can be exploited to 
construct compact three-qubit controlled gates using a small number of qubits. One example is shown in Fig. \ref{threespin1} {\bf (b)}, {where}
we require the logical qubits
to propagate away after the interaction via $\circlearrowright$ at the central triangle. It is easy to see that an enlargement of the basic three-spin triangle is necessary. In order to give a better insight into this, we have extracted the core entangled resource involved in simulating a Toffoli 
gate from Fig.~\ref{threespin1} {\bf (b)} into Fig.~\ref{threespin2} {\bf (a)}. The enlargement can be achieved by measuring qubits $7$ to $10$ in the $\sigma_x$ eigenbasis, a method similar to that used in the cluster-state model to remove pairs of adjacent qubits. The byproduct operation needed to retrieve the original ${\sf C^2P}$ of the central triangle between qubits $4,5$ and $6$ after enlargement ${\mathscr E}$ is given by $\tilde{U}_{\Sigma_{\mathscr E}}=\sigma_{z,4}^{s^x_8}\otimes\sigma_{z,5}^{s^x_9}\otimes\sigma_{z,6}^{s^x_7 s^x_{10}}{\sf CP}_{4,6}^{s^x_{10}}\otimes{\sf CP}_{5,6}^{s^x_{7}}$.
Here 
$s^x_i$ is the outcome of the measurement of qubit $i$ in the $\sigma_x$ eigenbasis with $s^x_i=0$ ($s^x_i=1$) corresponding to $\ket{+}_i$ ($\ket{-}_i=(1/\sqrt{2})[\ket{0}-\ket{1}]$). The 
${\sf CP}$'s in $\tilde{U}_{\Sigma_{\mathscr E}}$
result from
${\sf C^2P}$ not being in the Clifford group~\cite{RBH}.
In order to show that the enlarged three-spin triangle can be concatenated with the paths propagating the logical qubits toward and away from it, we write 
the {\sf CP} operator as
$S^{ij}=\openone^{(ij)}-2\ket{11}_{ij}\!\bra{11}$, which is
applied to qubits $i$ and $j$ in a triangle when the third qubit $k$ is in $\ket{1}$.
Let qubits $1,\,2$ and $3$ in Fig.~\ref{threespin1} {\bf (b)} encode the states $\ket{A},~\ket{B}$ and $\ket{C}$ respectively. Then, we entangle and measure the qubits as follows: $(\otimes^{3}_{i=1}P^{x}_{i})(\otimes^{3}_{1}S^{i,i+3})\ket{+}_{\cal T}\ket{A}_1\ket{B}_2\ket{C}_3$,
where $\ket{+}_{\cal T}=\otimes^{13}_{i=4}\ket{+}_i$ and $P^x_i$ represents the projector for a measurement in the $\sigma_{x,i}$ eigenbasis. This procedure realizes a two-{site cluster-state based propagation of the logical qubits}~\cite{RBH}, with qubits $7 \to 13$ left unaffected. Now, {let us} consider the {triangle enlargement}
described previously and {a subsequent information-flow away from it}. The entire process is written as 
$(\otimes^{10}_{i=1}P^x_{i})S^{6,13}S^{5,12}S^{4,11}S^{6,9,8}S^{5,10}S^{9,10}S^{7,8}S^{4,7}\times$
$(\otimes^{3}_{i=1}S^{i,i+3})\ket{+}_{\cal T}\ket{A}_1\ket{B}_2\ket{C}_3$.
As $[S^{ijk},\openone^{l}\!\otimes\!{S}^{mn}]=0$
($\forall{i,j,k,l,m,n}$) using the concatenation rules of propagation~\cite{RBH}, we see that the entire lattice can be entangled and then measurements performed. 
To complete this analysis,
we show how an arbitary byproduct operator changes on propagation through the ${\sf C^2P}$. Let $\otimes^{3}_{j=1}(\sigma_{x,\alpha_j}^{s^x_{\alpha_j}}\sigma_{z,\alpha_j}^{s^z_{\alpha_j}})$
denote the byproduct operator for any measurement pattern ${\cal M}$ carried out before the gate, where $\alpha_{j}$ is the site-label of the logical qubits, with values $s^x_{\alpha_j}$ and $s^z_{\alpha_j}$ dependent on the outcomes of ${\cal M}$ before $\alpha_j$. Upon propagation through, we obtain
 $\tilde{U}_{\Sigma_{\cal M}}=\prod_{j=1}^{3}({\cal P}_{j}{\sf CP}_{\alpha_2,\alpha_3}^{s^x_{\alpha_1}})$
$\times\prod_{j=1}^{3}({\cal P}_{j}\sigma_{x,\alpha_{1}}^{s^x_{\alpha_1}}\sigma_{z,\alpha_{1}}^{s^x_{\alpha_{2}}s^x_{\alpha_{3}}+s^z_{\alpha_{1}}})$, where ${\cal P}_{j}$ is the operator that exchanges label $1$ with $j$.
A two-qubit byproduct operator is again produced. For both $\tilde{U}_{\Sigma_{\cal M}}$ and $\tilde{U}_{\Sigma_{\mathscr E}}$, any ${\sf CP}$ cannot be propagated trivially and
it is necessary to remove it straight after the gate. This can be achieved by applying two-qubit gates analogous to the one in Fig.~\ref{threespin1} {\bf (a)} to logical qubits that underwent the $\sf C^2P$. In this case, we can measure the $b.q$'s in the $\sigma_y$ ($\sigma_z$)
eigenbasis, resulting in a ${\sf CP}$ (breaking the link) between the logical qubits up to local rotations \cite{Tame2}, thus reducing $\tilde{U}_{\Sigma_{{\cal M},{\mathscr E}}}$ to local forms again.
The simulation then proceeds as in the cluster-state model until the next three-qubit gate occurs.

{\it Physical Realization}- The trapping of alkali atoms such as $^{87}$Rb in hexagonal two-dimensional optical lattices can be achieved using three pairs of counter-propagating laser beams ($L,\,L^{\pm}$), tuned between the $D1$ and $D2$ line  with $\lambda=785~$nm and slightly detuned from each other. The pairs are in a lin$||$lin configuration~\cite{Cal1} and propagate along $\hat{\bf y}$ and $(\hat{\bf y}\pm\sqrt{3}\hat{\bf x})/2$ respectively, providing lattice sites with periodicity $\lambda / \sqrt{3}$. An appropriate external trapping field is applied in the $\hat{\bf z}$ direction to confine the atoms to the ${x}-{y}$ plane. Each logical qubit can be embodied by the single-atom hyperfine states $\ket{a}=\ket{0}\equiv\ket{F=1, m_f=1}$ and $\ket{b}=\ket{1}\equiv\ket{F=2, m_f=2}$, with $F$ and $m_{f}$ the total angular momentum of the atom and its projection along ${\bf \hat{z}}$ respectively.
They can then be coupled via a Raman transition~\cite{Jak1}, using 
an excited state $\ket{e}$ embodied by another hyperfine state.
We assume the lattice is initially loaded with one atom per site,
which can be achieved by making a Bose-Einstein condensate undergo a superfluid to Mott insulator (MI) phase transition~\cite{Jak1, Mandel1,W}. In a second-quantization picture, the two-species Bose-Hubbard Hamiltonian $H=-\sum_{\alpha=a,b}(J^{\alpha}\sum_{\langle{i,j}\rangle}\alpha_i^{\dag}\alpha_j-\frac{1}{2}U^{\alpha\alpha}\sum_{i}\alpha_i^{\dag 2}\alpha_i^2)+U^{ab}\sum_{i}a_i^{\dag}b_i^{\dag}a_i b_i$ describes the dynamics of the cold gas of interacting bosons
in the periodic trapping potential considered.
Here, $\alpha_i$ ($\alpha^{\dag}_i$) [$\alpha=a,b$] are the annihilation (creation) operators for atomic species $\alpha$ at site $i$ and the summations are taken over nearest-neighbor (NN) sites, indicated by $\langle{i},\,j\rangle$. 
$U^{\alpha\alpha}$ $(U^{ab}=U^{ba})$ represents the strength of the homo-species (hetero-species) on-site repulsive force experienced by the atoms.
$J^{\alpha}$ is the homo-species tunneling strength between sites $i$ and $j$.
For both $U$ and $J$, effects from atoms in NN and next-NN sites respectively are assumed to be negligible~\cite{Jak1}.
For a hexagonal periodic potential, we can confine the Hamiltonian to a unit-cell of three sites in an equilateral triangular configuration~\cite{Pachos1} obtaining $H=H^{(0)}+V$, where
$H^{(0)}= \frac{1}{2}\sum_{i\,\alpha\,\beta}U^{\alpha \beta} \alpha_i^{\dag} \beta_i^{\dag} \beta_i \alpha_i$ and  
$V=-\sum_{i\,\alpha} J_{i}^{\alpha} (\alpha_i^{\dag}\alpha_{i+1}+\alpha_{i+1}^{\dag}\alpha_{i} )$.
For {$J^{\alpha}_{j}\ll{U}^{\alpha\,\beta}\,(\forall{j}$), $\alpha,\beta\in\{a,b\}$, $V$ can be treated as a perturbation. In these conditions, 
leaving the subspace $M$ corresponding to the prepared MI (with unit filling-fraction) is energetically unfavorable for the system: its spectrum is gapped and states with higher filling-fractions can be 
adiabatically eliminated using standard techniques~\cite{Pachos2}. An effective Hamiltonian is obtained with only virtual transitions to higher population subspaces~\cite{Pachos1}.
By switching to the pseudo-spin basis
$\vert{\uparrow}\rangle=\ket{n_{i}^a=1,\,n_{i}^b=0}$
and $\vert{\downarrow}\rangle=\ket{n_{i}^a=0,\,n_{i}^b=1}$, this Hamiltonian
can be written as
~\cite{Pachos2} $H_{{eff}}=\sum_{j=1}^3[A_{j}\openone+\lambda^{(0)}_{j}
\sigma^z_j+\lambda^{(1)}_{j}H^{I}_{j,j+1}
 +\lambda^{(2)}_{j}H^{XY}_{j,j+1}+\lambda^{(3)}H^{T}_{j,j+1,j+2}+\lambda^{(4)}_{j}H^{XYZ}_{j,j+1,j+2}]$, where 
$H^{I}_{j,j+1}=\sigma^z_j \sigma^z_{j+1}$, 
$H^{XY}_{j,j+1}=\sum_{l=x,y}\sigma^l_{j}\sigma^l_{j+1}$,
$H^{T}_{j,j+1,j+2}=\sigma^z_{j} \sigma^z_{j+1}
  \sigma^z_{j+2}$ and  $H^{XYZ}_{j,j+1,j+2}=\sum_{l=x,y}\sigma^l_{j} \sigma^z_{j+1}\sigma^l_{j+2}$.
Here $A$ and the $\lambda^{(i)}$'s depend on $J^{\alpha}_{j}$ and $U^{\alpha\beta}$. By varying the laser parameters, 
specific parts of $H_{{eff}}$ can dominate over the remainder~\cite{Pachos3}. In particular, we are interested in the terms containing $\lambda^{(0)}_j,\,\lambda^{(1)}_j$ and $\lambda^{(3)}$.
Later, we address a suitable choice for the physical parameters in $H_{eff}$. 
To realize a ${\sf C^2P}$ from $H_{{eff}}$, 
$\Lambda_{0}=\Lambda_{3}=-\Lambda_{1}=\pi/8$ is required, where $\Lambda_{i}=\int_{0}^{T} \lambda^{(i)} dt$. This is possible by tuning $J^{\alpha}$, $U^{\alpha\beta}$ and $T$.
However a restriction is imposed
by the condition $U/zJ \gtrsim 5.8$, which guarantees the MI regime with one atom per site~\cite{mft},
 where $z$ is the number of NN
seen by a given site. 
$H_{{eff}}$ can be generalized to the hexagonal lattice, with each site having 6 NN, so that the model is valid for $J/U \lesssim 0.03$. 

In order to produce the bowtie pattern, we can
create an optical potential of period $>\lambda /2$~\cite{Peil}. Here, two laser-beams, 
at angles $\pm\theta/2$ to a given direction $\vec{v}$ on the $x-y$ plane
produce
a two-dimensional standing-wave in the direction perpendicular to $\vec{v}$ on the $x-y$ plane with period {\bf $d=\lambda /[2 \sin(\theta/2)]$}. Using two pairs of lasers, we can produce the periodic pattern $V_{off}=V_0 - V_1 \cos(|\vec{k}| \hat{\bf y})+ V_2 \cos(|\vec{k}| \sqrt{3} \hat{\bf x})$ that offsets the original hexagonal lattice as shown in Fig.~\ref{threespin2} {\bf (b)}.
Here $V_1$ ($V_2$) is a potential produced by the first (second) pair of lasers, $V_0=V_{1}+V_{2}$
and $|\vec{k}|= 2 \pi /\lambda$. $V_{off}$ suppresses tunneling
between specific sites on the lattice according to the pattern in Fig.~\ref{threespin2} {\bf (b)}.
During the time evolution, a ${\sf C^2P}$ is realized
between the sites of closed triangles only.

The initialization of the register prior to the entanglement
is achieved by applying 
Raman transitions to all lattice sites. These can be activated by standing-waves of period $a$ from two pairs of lasers $L_1$ and $L_2$, far blue-detuned by $\Delta$ from the transition $\ket{\{a,b\}}\leftrightarrow\ket{e}$ and orientated along the ${\bf \hat{y}}$-axis. All the sites will be located at the maximum-intensity peaks~\cite{Pachos4} and with the atoms initially in $\ket{a}$, 
a rotation of the qubits into the state $\ket{+}$ can be achieved.
In order to perform information-flow as described above,
individual qubits along the edges of a path must be set to $\ket{\{0,1\}}$. It is experimentally feasible to apply a Raman transition to a bunch of qubits by addressing them with two lasers of cross-section $\sigma\simeq{10^{-12}}m^{2}$. With a Gaussian radial intensity-profile, positioning the beams' center between the atoms to be addressed, as shown in Fig.~\ref{threespin2} {\bf (b)}, applies the same transition to all the closest surrounding atoms. Thus the qubits may be rotated from $\ket{+}$ to $\ket{0}$ or $\ket{1}$ as needed using a sort of {\it blurred removal}~\footnote{Alternatively, a far-off-resonant laser can be focused to a fraction of the lattice spacing, Stark shifting $\ket{a}$ and $\ket{b}$ differently. A global addressing microwave pulse then only rotates states of atoms that are exactly in resonance.}. The state of the center qubit is irrelevant, as it will be disentangled from the rest of the lattice. This allows the overlapping of blurred removals and the creation of entangled-state subspaces separated from the rest of the lattice. Other techniques, such as using a diffraction limited lens system~\cite{Baran}, would also be suitable for this task.
Such methods can also be used in preparing cluster-states in square lattice configurations, although single-atom addressing would obviously allow more compact gate constructions in both models. 

{\it Imperfect operation}-
The realization of a single lattice-wide ${\sf C^2 P}$
should be carried out within the system's coherence time. 
Both $\lambda^{(1)}$ and $\lambda^{(3)}$ 
roughly scale as $J^3/U^2$ ($J^{\alpha}_i=J^{\alpha},~\forall i$), with variability upper-bounded by $J/U \sim 0.03$. 
One can set $J$ and $U$ so that $\vert\lambda^{(0,1,3)}\vert\sim 0.3$ Hz and $\lambda^{(2,4)}=0$ ($\lambda^{(0)}$ can be adjusted by an appropriate Zeeman term~\cite{Pachos3}). 
Here we take $J^b\simeq 0$, $J^a=2$ kHz and $U^{\alpha\alpha}=2U^{\alpha\beta}=120$ kHz, possible using Feshbach resonances~\cite{Vogels},
which correspond to interaction-times within the coherence time of this far-detuned configuration~\cite{dec}. Couplings one order of magnitude larger are possible, thus lowering the operation time, with the requirement that $J$ and $U$ increase by an order of magnitude. Small deviations from the desired values of $J$ and $U$ 
imply slight variations of $\lambda^{(i)}$'s 
with $\lambda^{(2,4)}$ 
becoming nonzero and affecting the system.
In general, as $\lambda^{(2)}\simeq10\lambda^{(4)}$
for the choices above, 
we neglect its effect. Thus, 
the replacements 
$\lambda^{(j)}\rightarrow \lambda^{(j)}+ \epsilon_j\,(j=0,..,3)$, 
in $H_{eff}$ allow for the formal study of imperfect entanglement-generation.
The imperfect Hamiltonian evolves the initial state $\ket{\psi(0)}=\sum^{1}_{\beta,\gamma,\delta=0}\alpha_{\beta\gamma\delta}(0)\ket{\beta,\gamma,\delta}$ of qubits within a triangle  
into the state 
$\ket{\psi(t)}$. To determine the quality of the dynamics, we compare $\ket{\psi(t)}$ to $\ket{\psi^I}={\sf C^2 P}\ket{\psi(0)}$ using the fidelity
$F=|\braket{\psi^I}{\psi(t)}|^2$. 
\begin{figure}[t]
{\bf (a)}\hskip4cm{\bf (b)}
\centerline{\psfig{figure=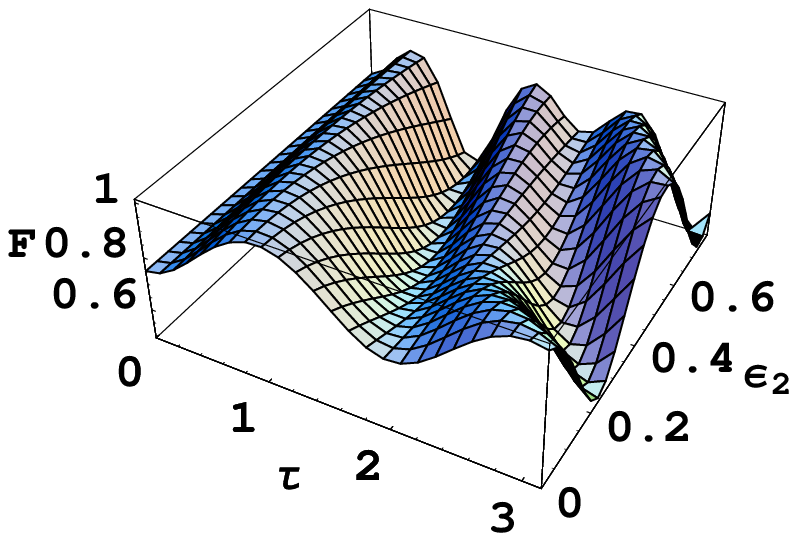,width=4.3cm,height=3.4cm}\psfig{figure=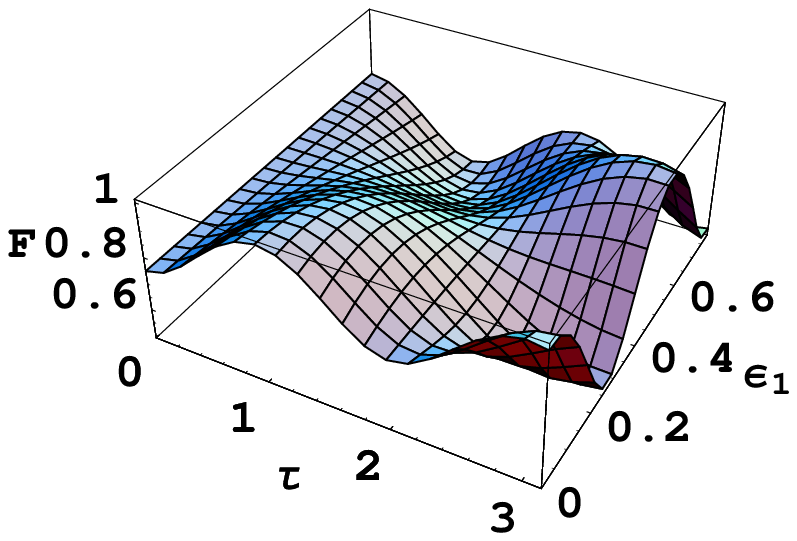,width=4.3cm,height=3.4cm}}
\caption{{\bf(a)}: The fidelity $F$ of the entanglement generation against the rescaled time $\tau$ and the coupling $\epsilon_2$. 
{\bf (b)}: Similar to {\bf (a)} but with the replacement $\epsilon_{2}\rightarrow\epsilon_1$, to study imperfectness in the $H^{I}$ coupling.} 
\label{fidelity}
\end{figure}
Fig.~\ref{fidelity} {\bf (a)} shows the result for $\alpha_{i}(0)=1/2\sqrt{2}$.  
Here $\epsilon_2$ has been allowed to vary with $\epsilon_{0,1,3}=0$. For convenience, we have set $\vert\Lambda_{0,1,3}\vert=(\pi/8)\tau$.
For $\tau=1$, corresponding to an application of $\sf C^2P$ if $\epsilon_{2}=0$, there is a noticeable fidelity decay against $\epsilon_{2}$. Moreover, for $\epsilon_{2}>0$, $F=1$ at $\tau<{1}$, which may allow a compensation for the fidelity decay 
by using a shorter interaction-time. However this procedure is ineffective if
$\epsilon_2$ is unknown. Fig.~\ref{fidelity} {\bf (b)} shows another example of the effect of imperfect couplings, where we have 
allowed $\epsilon_1$ to vary, keeping $\epsilon_{0,2,3}=0$, a choice due to $\lambda^{(1)}$ changing more rapidly than $\lambda^{(3)}$. It is clear that the parameters in $H_{eff}$ should be accurately tuned if the correct entanglement is to be realized.

{\it Remarks}- We have shown that one-way \QC~can be performed on a three-body based 
entangled resource.
Using concatenation methods we have described an economic Toffoli gate simulation, fundamental to compact algorithm realizations and  robust-to-noise QIP.
We have also analyzed in detail the feasibility of an optical lattice-based implementation of our model.

{\it Acknowledgements}- We thank 
P. B. Blakie and C. J. Foot for discussions. This work has been supported by EPSRC, DEL and KRF.



\begin{thebibliography}{99}

\bibitem{RH} R. Raussendorf and H. J. Briegel, {\sl Phys. Rev. Lett.} {\bf 86}, 5188 (2001).

\bibitem{RBH} R. Raussendorf, D. E. Browne, and H. J. Briegel, {\sl Phys. Rev. A} {\bf 68}, 022312 (2003).

\bibitem{HEB} M. Hein, J. Eisert, and H. J. Briegel, {\sl Phys. Rev. A} {\bf 69}, 062311 (2004).


\bibitem{DH} W. D\"ur and H. J. Briegel, {\sl Phys. Rev. Lett.} {\bf 92}, 180403 (2004).

\bibitem{Wal} P. Walther {\it et al.}, {\sl Nature} (London) {\bf 434}, 169 (2005).

\bibitem{Pan} A.-N. Zhang {\it et al.}, {\sl quant-ph/0501036} (2005).

\bibitem{Tame1} M. S. Tame, M. Paternostro, M. S. Kim, and V. Vedral, {\sl quant-ph/0412156} (2004).

\bibitem{Tame2} M. S. Tame, M. Paternostro, M. S. Kim, and V. Vedral, {\sl quant-ph/0502081} (accepted in Phys. Rev. A) (2005).


\bibitem{Bar} A. Barenco {\it et al.}, {\sl Phys. Rev. A} {\bf 52}, 3457 (1995).

\bibitem{Grov} L. Grover, {\sl Phys. Rev. Lett.} {\bf 79}, 325 (1997).

\bibitem{Conc} M. S. Tame, unpublished. A concatenation of Toffoli gates rather than stabilizer formalism is assumed~\cite{Tame2}.

\bibitem{Pachos1} J. K. Pachos and M. B. Plenio, {\sl Phys. Rev. Lett.} {\bf 93}, 056402 (2004).

\bibitem{Cal1} T. Calarco {\it et al.}, {\sl J. Mod. Opt.} {\bf 47}, 2137 (2000).

\bibitem{Jak1} D. Jaksch {\it et al.}, {\sl Phys. Rev. Lett.} {\bf 81}, 3108 (1998).

\bibitem{Mandel1} O. Mandel {\it et al.}, {\sl Nature} (London) {\bf 425}, 937 (2003).

\bibitem{W} Inhomogeneities in the atomic distribution can be fixed as in 
P. Rabl {\it et al.}, {\sl Phys. Rev. Lett.} {\bf 91}, 110403 (2003), D. S. Weiss {\it et al.}, {\sl Phys. Rev. A} {\bf 70}, 040302(R) (2004).

\bibitem{Pachos2} J. K. Pachos and E. Rico, {\sl Phys. Rev. A} {\bf 70}, 053620 (2004). 


\bibitem{Pachos3} J. K. Pachos {\it et al.}, {\sl Opt. Spectrosc.} {\bf 99}, 355 (2005).

\bibitem{mft} M. P. A. Fisher {\it et al.}, {\sl Phys. Rev. B} {\bf 40}, 546 (1989).

\bibitem{Peil} S. Peil {\it et al.}, {\sl Phys. Rev. A} {\bf 67}, 051603 (2003).

\bibitem{Pachos4} A. Kay and J. K. Pachos, {\sl New J. Phys.} {\bf 6}, 126 (2004).


\bibitem{Baran} P. Baranowski, J. Zacks, G. Hechenblaikner and C. J. Foot, {\sl physics/0412126} (2004).

\bibitem{Vogels} J. M. Vogels {\it et al.}, {\sl Phys. Rev. A.} {\bf 56}, 1050 (1997).

\bibitem{dec} Q. Thommen, J. C. Garreau, and V. Zehnl$\acute{\rm e}$, {\sl Am. J. Phys.} {\bf 72}, 1017 (2004).

\end{thebibliography}
\end{document}